\documentclass{article}

\PassOptionsToPackage{numbers, compress}{natbib}

\usepackage{graphicx}
\usepackage{subcaption}

\usepackage[final]{neurips_2022}

\usepackage[utf8]{inputenc} 
\usepackage[T1]{fontenc}    
\usepackage{hyperref}       
\usepackage{url}            
\usepackage{booktabs}       
\usepackage{amsfonts}       
\usepackage{nicefrac}       
\usepackage{microtype}      
\usepackage{xcolor}         

\title{Visualizing DNA reaction trajectories with deep graph embedding approaches}

\author{
  Chenwei Zhang \\
  Department of Computer Science\\
  University of British Columbia\\
  Vancouver, BC V6T1Z4 \\
  \texttt{cwzhang@cs.ubc.ca} \\
  \And
  Khanh Dao Duc\\
  Department of Mathematics \\
  University of British Columbia\\
  Vancouver, BC V6T1Z4 \\
  \texttt{kdd@math.ubc.ca} \\
  \And
  Anne Condon \\
  Department of Computer Science \\
  University of British Columbia\\
  Vancouver, BC V6T1Z4 \\
  \texttt{condon@cs.ubc.ca} \\
}

\begin{document}

\maketitle

\begin{abstract}
Synthetic biologists and molecular programmers design novel nucleic acid reactions, with many potential applications. Good visualization tools are needed to help domain experts make sense of the complex outputs of folding pathway simulations of such reactions. Here we present ViDa, a new approach for visualizing DNA reaction folding trajectories over the energy landscape of secondary structures. We integrate a deep graph embedding model with common dimensionality reduction approaches, to map high-dimensional data onto 2D Euclidean space. We assess ViDa on two well-studied and contrasting DNA hybridization reactions. Our preliminary results suggest that ViDa's visualization successfully separates trajectories with different folding mechanisms, thereby providing useful insight to users, and is a big improvement over the current state-of-the-art in DNA kinetics visualization.
\end{abstract}

\section{Introduction}
Nucleic acid nanotechnologies, including DNA beacons~\cite{Wang2009-nk}, RNA riboswitches ~\cite{Roth2009-tb}, as well as DNA implementations of Boolean circuits~\cite{qian-winfree-2011} and artificial neural networks~\cite{cherry-qian-2018}, employ multi-stranded reactions such as DNA hybridization, three-way toehold-mediated strand displacement, and four-way branch migration. Molecular programmers are keen to better understand the rates of such reactions since, even for strands of the same length, rates can vary dramatically as a function of sequence. However, the mechanisms that determine nucleic acid reaction kinetics are not well understood. Emerging geometric deep learning methods \cite{graphconv,diffusioncnn,GSAE} have demonstrated success in analyzing graph-based data, providing a new strategy to represent DNA and RNA secondary structures and energy landscapes. Inspired by this, we propose \textbf{ViDa}, a new visualization approach that integrates recent deep graph embedding and dimensionality reduction methods to help researchers learn about DNA reaction mechanisms from computationally sampled secondary structure folding trajectories. We present a case study that shows the effectiveness of ViDa, and also reveals strengths and weaknesses of the ML tools employed in ViDa, helping to guide future work on ML model development.

\subsection{Background and related work} \label{relatedwork}
To analyze reaction mechanisms, we are interested in studying trajectories, i.e., sequences of secondary structures, from the reactants to the products of a DNA reaction, along with the time to transition from one secondary structure to the next.
A secondary structure describes the set of base pairs formed via hydrogen bonding between Watson-Crick complementary bases, and each secondary structure has an associated free energy that is determined by latent thermodynamic parameters. Elementary step simulators such as Multistrand~\cite{multistrand} (see Appendix \ref{multistrand}) implicitly use continuous-time Markov chain models of reaction trajectories, and can stochastically generate trajectory samples. Multistrand's output just uses “dot-parenthesis” (dp) notation (see Appendix \ref{dp} and examples in Table \ref{table:samples}) to represent a secondary structure, and a sequence of such strings to represent a trajectory. This output does not situate the trajectory in the overall energy landscape. Machineck et al. \cite{MachinekThreeway} used a coarse-grained map to show reaction trajectories laid on an energy landscape. However, the coarse-grained grid cells may include secondary structures with very different free energies, making interpretation of different reaction trajectories difficult.

Castro et al. \cite{GSAE} developed a deep graph embedding framework, called the geometric scattering autoencoder (GSAE) network to study RNA secondary structure landscapes. GSAE has three major parts: an untrained geometric scattering network \cite{scatteringtransform}, a trained variational autoencoder (VAE) \cite{VAE}, and a trained auxiliary regression network, where the latter two networks assemble to form as a semi-supervised variational autoencoder. The geometric scattering transform ﬁrst extracts features, called scattering coefﬁcients, from the input graph. These high-dimensional coefﬁcients are embedded into low-dimensional representations through the trained semi-supervised VAE network in light of their secondary structures and metaproperties, such as energy, concentration, and so on. The embedded features retain necessary information that can be used for further study. However, this approach is currently limited to single-stranded secondary structures, whereas important nucleic acid reactions are typically multi-stranded, and does not address how to visualize trajectories through those landscapes.

Although the dimension of the vector output by VAE is smaller than that produced by the geometric scattering transform, it needs to be reduced further. We compare several dimensionality reduction approaches that further reduce the number of features in the input data, while retaining important information. In this work we apply PCA~\cite{PCA}, a linear method, as well as the non-linear methods including PHATE \cite{PHATE}, t-SNE \cite{tSNE}, and UMAP \cite{UMAP}. 

\subsection{Primary contributions}
We introduce a new workflow for dimensionality reduction and visualization approach, called ViDa, based on deep graph embedding, for visualizing and exploring DNA reaction trajectories. We evaluate ViDa on two well-studied DNA hybridization reactions~\cite{gaohelix} with contrasting reaction folding pathways (see details in Section \ref{evl}). Our assessment is motivated by two questions: (i) Do ViDa's energy landscape embeddings preserve local and global structure? That is, are secondary structures with similar base pairs and energies clustered together (local preservation), and does the distance between clusters reflect dissimilarities between their secondary structures (global preservation)? (ii) More importantly, do trajectories laid out on the energy landscapes reveal meaningful differences in folding pathways between the two reactions? Our results demonstrate that ViDa can help domain experts gain insight from sampled reaction trajectories, because of its good interpretability. Furthermore, the results also suggest that ViDa can help evaluate the performance of the employed ML tools, guiding better development of ML approaches.

\section{Methods} \label{headings}

\begin{figure}[h]
  \centering
  \includegraphics[width=1\linewidth]{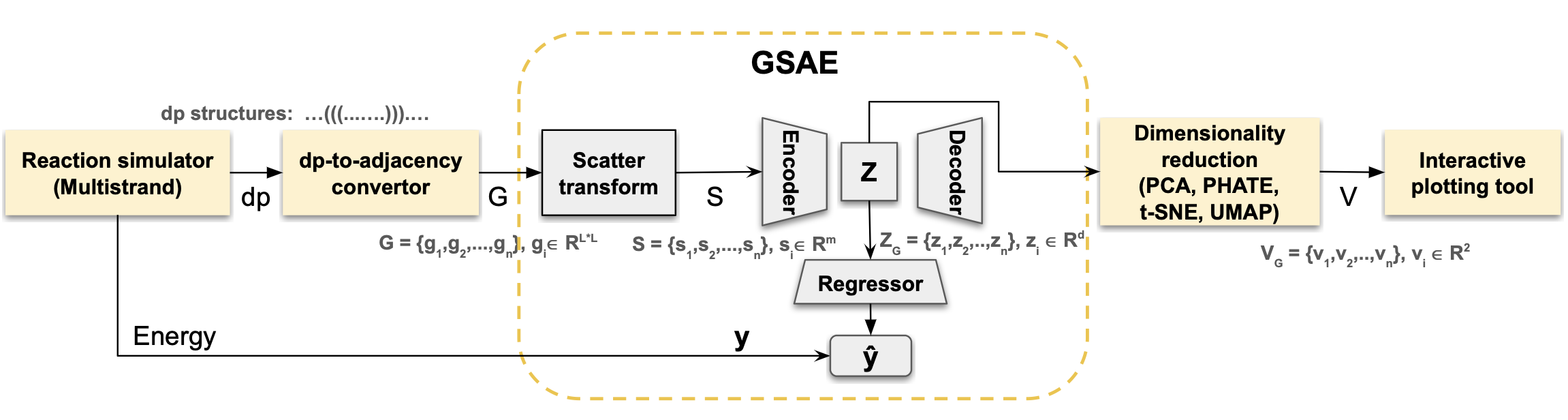}
  \caption{The framework of ViDa. The scattering transform and semi-VAE make up the GSAE model. $\hat{y}$ is the predicted energy by the regressor network and y is the real energy produced by the simulator.}
  \label{fig:ViDa}
\end{figure}

\subsection{ViDa workflow}
The ViDa framework pipeline, illustrated in Figure \ref{fig:ViDa}, consists of five major parts: the Multistrand reaction simulator~\cite{multistrand}, a converter which produces adjacency matrices from dp notation, a graph embedding model (GSAE), a dimensionality reduction technique (PCA, PHATE, t-SNE, or UMAP), and an interactive plotting tool. An input set of secondary structures, represented using dp notation, and their corresponding energies, were extracted from sampled Multistrand trajectories. Each secondary structure was converted to a graph adjacency matrix, with one node per nucleotide and two types of edges: those that represent complementary base pairs in the structure, and those representing the strand backbones. The resulting set of graphs $G=\{g_1,g_2,...,g_n\}, g_i \in \mathbb{R}^{L \times L}$ (where $L=50$ is the sum of the lengths of two single-stranded sequences) were input to a geometric scattering transform network, which converts graph signals to scattering coefficient vectors. 70\% of coefficient vectors were randomly assigned to the training set for the VAE model, and the remaining 30\% were assigned to the testing set. The latent embedding $Z$ that was produced by the VAE encoder was input to the auxiliary regression network, which penalizes incorrect energy predictions. Once training completed, we fed all converted scattering coefficient vectors to the trained VAE model which maps them to a set of vectors $Z_G=\{z_1,z_2,...,z_n\}, z_i \in \mathbb{R}^d$ in d-dimensional Euclidean space. Finally, the dataset $Z_G$ was input to dimensionality reduction algorithms (PCA, PHATE, t-SNE, or UMAP) to obtain an embedding $V_G=\{v_1,v_2,..,v_n\}, v_i \in \mathbb{R}^2$ in 2D Euclidean space for visualization purposes.

\subsection{Case study design} \label{evl}
To evaluate ViDa, we use two DNA hybridization reactions that have been studied experimentally by Gao et al. \cite{gaohelix} (see also a computational study by Schreck et al. \cite{schreck}). For one reaction, bases of a strand were designed to easily ``zip up'' with the bases of the complementary strand to form a duplex. In contrast, in the second reaction, the complementary strands were designed so that intra-strand hairpin formation could impede the zippering process and slow down duplex formation. See Table \ref{table:samples}. We generated samples of reaction folding trajectories using the Multistrand simulator. For each reaction, the set of secondary structures visited in the sampled trajectories comprises the landscape.

\subsection{Implementation}
The VAE model was trained separately for the P0+T0 sample and the P4+T4 sample. The total loss of the model is made up of three terms: two from the VAE model, namely reconstruction and latent losses; and one from the auxiliary regression network, regression loss. For the experimental results presented, the iteration epochs were 60 for the P0+T0 sample and 100 for the P4+T4 sample, to avoid overﬁtting with a batch size of 64. The bottleneck dimension of the VAE was set to 25. The optimization was done by PyTorch’s Adam optimizer with a learning rate of 0.0001. The hyperparameters $\alpha$ and $\beta$ were set to 1 and 0.0001 respectively, to control the importance of the regression and latent losses.

Using PHATE, the number of landmarks was set to 2000. The decay rate was set to 40 and the number of nearest neighbours was set to 5. 
The parameter settings for t-SNE and UMAP are provided in Appendix \ref{tsneumap}. The interactive plotting tool used the Plotly library in Python.
\footnote{Our code is available at the GitHub repository \url{https://github.com/chenwei-zhang/ViDa}.}

\begin{table}[h]
  \caption{Data samples P0+T0 and P4+T4, each with two complementary stands~\cite{gaohelix}. Initial structures for both samples are unpaired single-stranded secondary structures, while final structures have all bases paired. Strands P0 and T0 were designed so that no stable intra-strand hairpins can form, while P4 and T4 were designed so that a stable hairpin with four paired bases can form. Schematic representations for hairpin structures are shown in Figure \ref{fig:hairpins}.\\}

  \label{table:samples}
  \small
  \centering
  \begin{tabular}{ll}
    \midrule
    Sample P0+T0: & 3$'$-\texttt{GAGACTTGCCATCGTAGAACTGTTG}-5$'$ + 3$'$-\texttt{CAACAGTTCTACGATGGCAAGTCTC}-5$'$ \\
      Initial structure: & 3$'$-\texttt{.........................}-5$'$ + 3$'$-\texttt{.........................}-5$'$ \\
    Final structure: & 3$'$-\texttt{(((((((((((((((((((((((((}-5$'$ + 3$'$-\texttt{)))))))))))))))))))))))))}-5$'$ \\
    \midrule
    Sample P4+T4: & 3$'$-\texttt{ACACGATCATGTCTGCGTGACTAGA}-5$'$ + 3$'$-\texttt{TCTAGTCACGCAGACATGATCGTGT}-5$'$  \\
      Initial structure: & 3$'$-\texttt{.........................}-5$'$ + 3$'$-\texttt{.........................}-5$'$ \\
    Final structure: & 3$'$-\texttt{(((((((((((((((((((((((((}-5$'$ + 3$'$-\texttt{)))))))))))))))))))))))))}-5$'$ \\
    4-stem hairpins: & 3$'$-\texttt{.((((..........))))......}-5$'$ + 3$'$-\texttt{.....((((...........)))).}-5$'$ \\
    \bottomrule
  \end{tabular}
\end{table}

\section{Results} \label{results}

We present here the results obtained from running ViDa with PCA and PHATE. For a discussion of the results obtained with t-SNE and UMAP, see Section \ref{suggestion} and Appendix \ref{tsneumap}.

\begin{figure}[h]
    \centering
    \begin{subfigure}[b]{0.33\textwidth}
        \includegraphics[width=1\linewidth, trim={1cm 0cm 4.5cm 6cm} ,clip] 
        {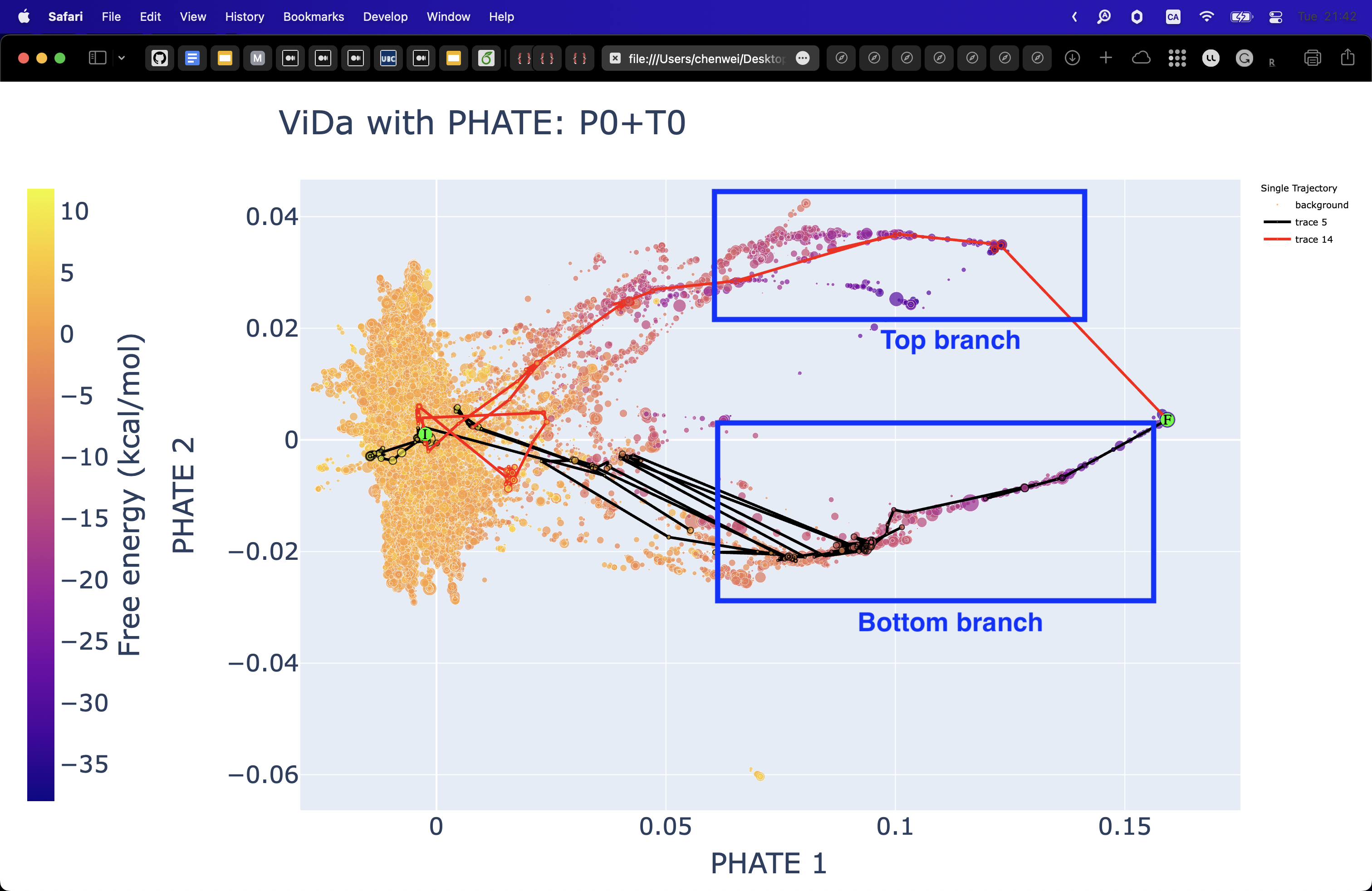}\hfil
        \caption{ViDa with PHATE: P0+T0 \label{fig:result-a}}
    \end{subfigure}
    \begin{subfigure}[b]{0.33\textwidth}
        \includegraphics[width=0.905\linewidth, trim={5.8cm 0cm 4.5cm 6cm} ,clip]
        {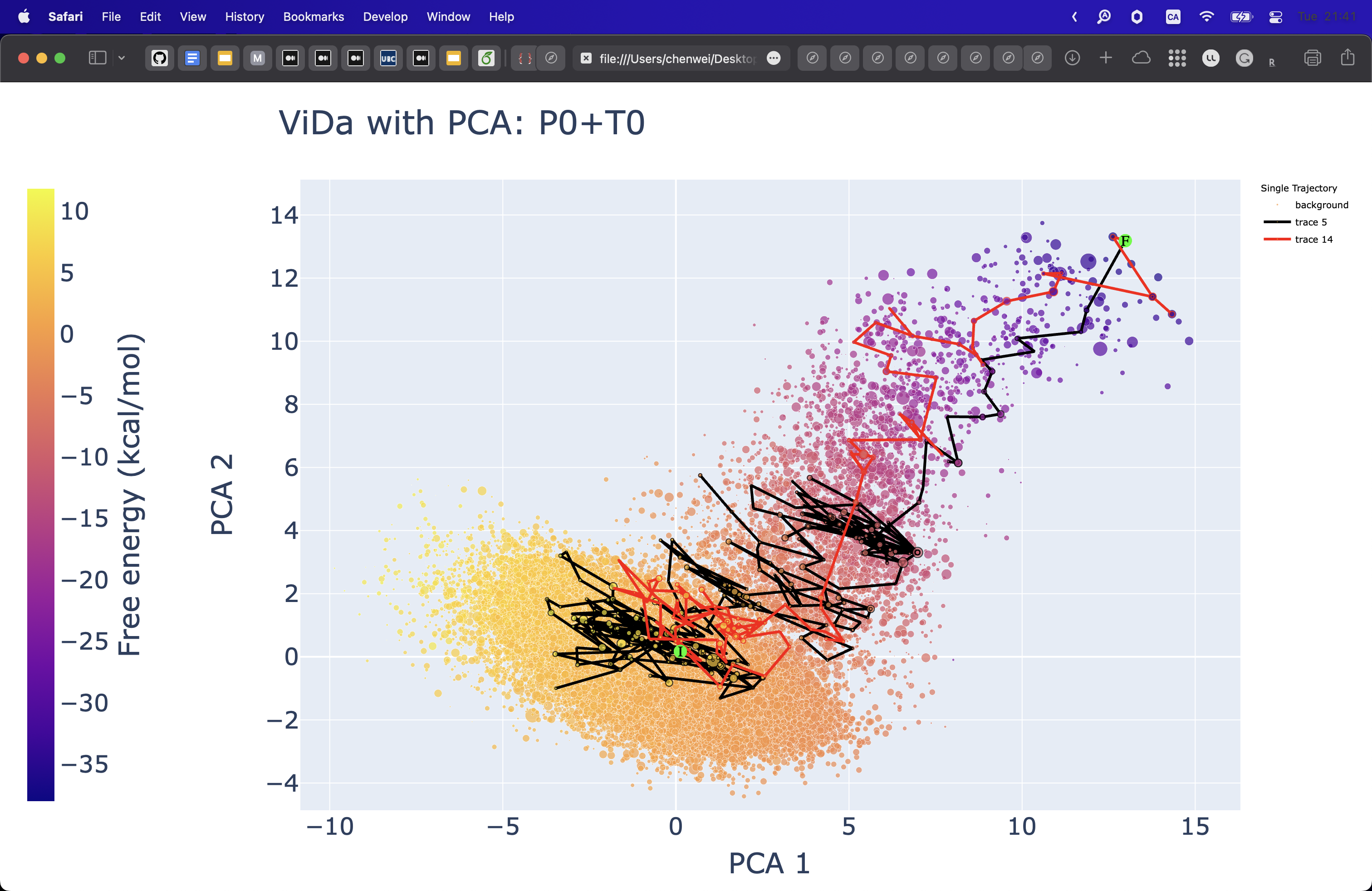}
        \caption{ViDa with PCA: P0+T0 \label{fig:result-b}}
    \end{subfigure}\hfil
    \begin{subfigure}[b]{0.33\textwidth}
        \includegraphics[width=1\linewidth, trim={1cm 0cm 4.5cm 6cm} ,clip] 
        {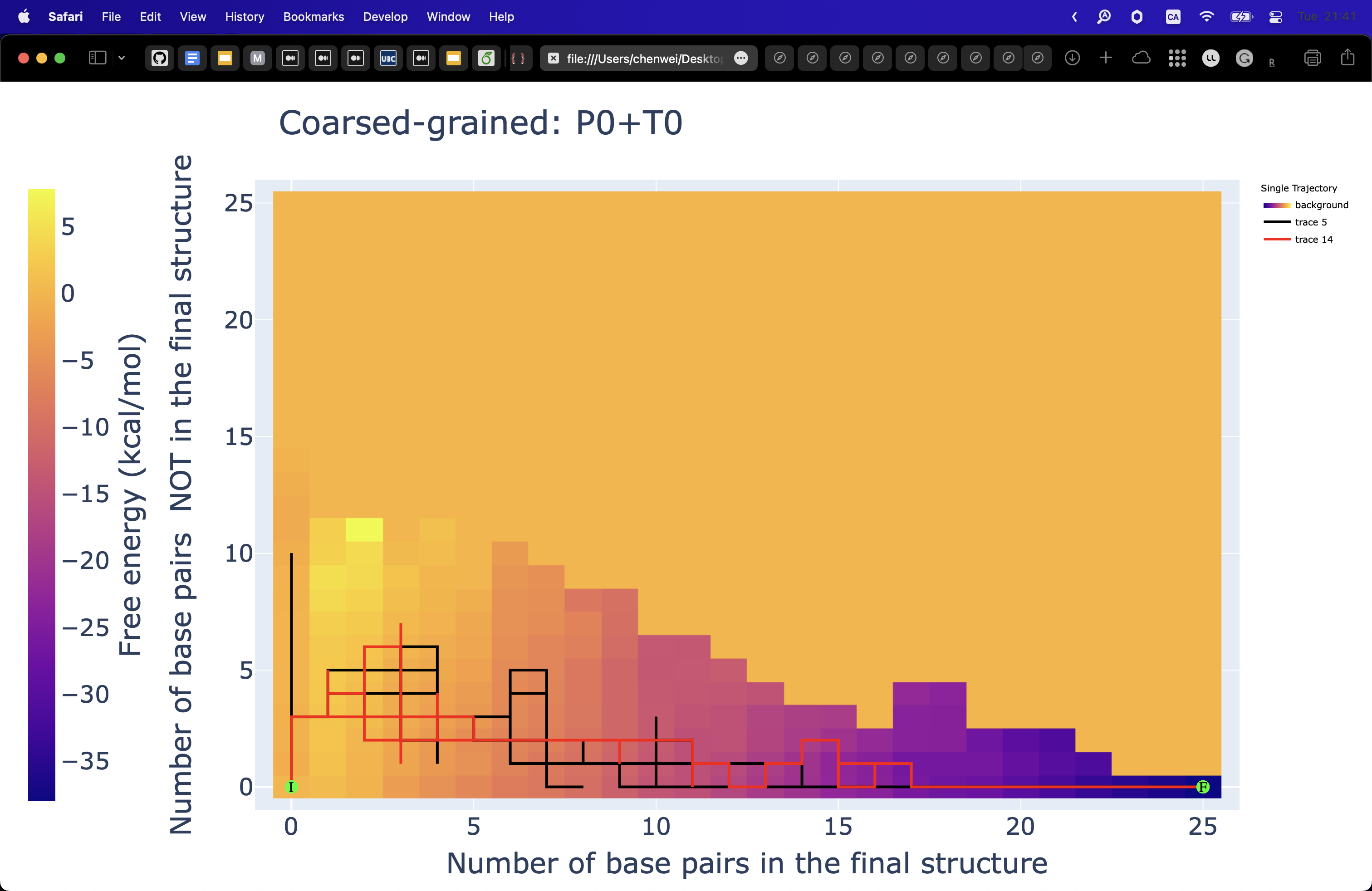}\hfil
        \caption{Coarse-grained: P0+T0 \label{fig:result-cg0}}
    \end{subfigure}\hfil
    \begin{subfigure}[b]{0.33\textwidth}
        \includegraphics[width=1\linewidth, trim={1cm 0cm 4.5cm 6cm} ,clip] 
        {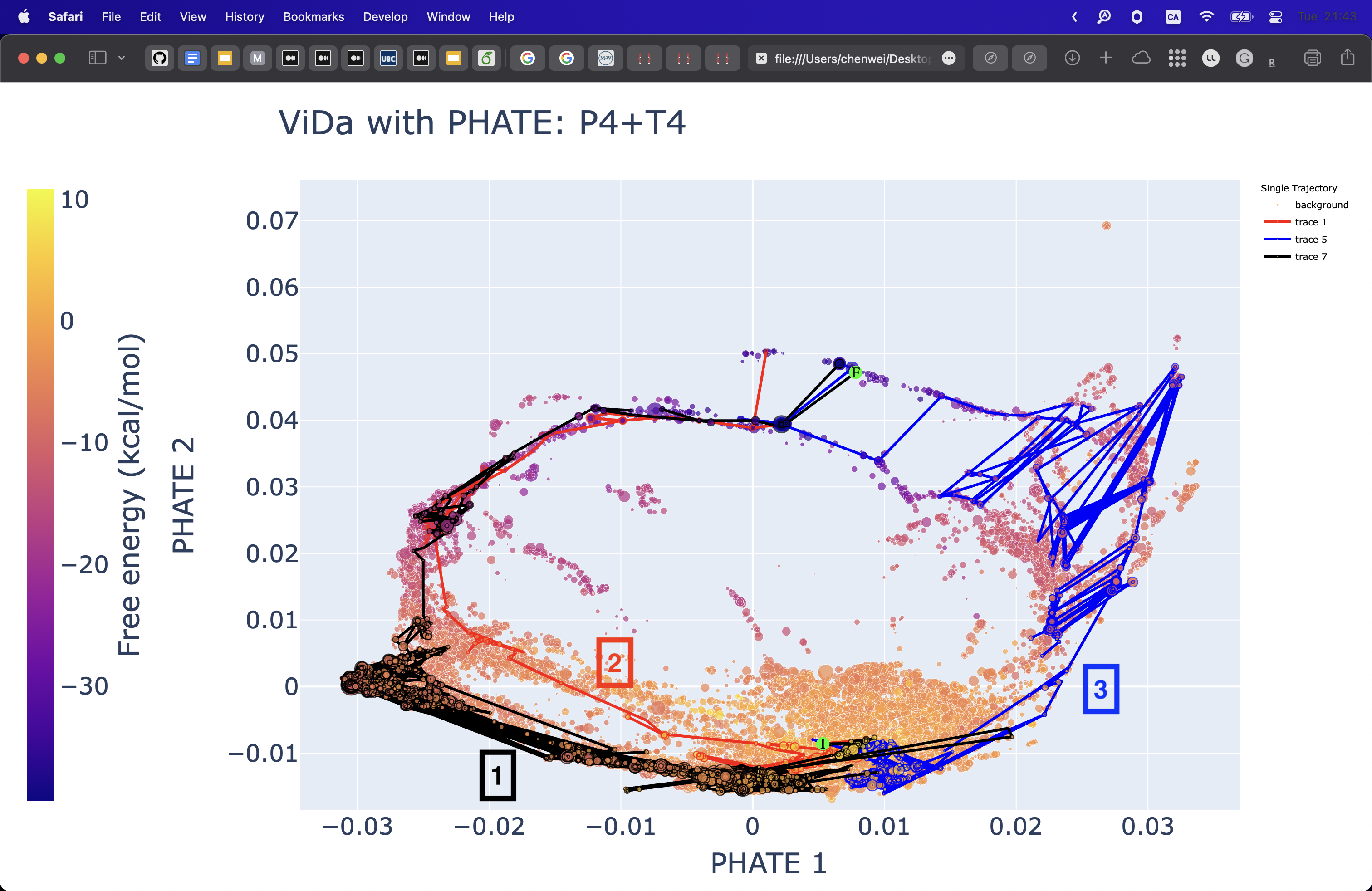}\hfil
        \caption{ViDa with PHATE: P4+T4 \label{fig:result-c}}
    \end{subfigure}
    \begin{subfigure}[b]{0.33\textwidth}
        \includegraphics[width=0.905\linewidth, trim={5.8cm 0cm 4.5cm 6cm} ,clip]
        {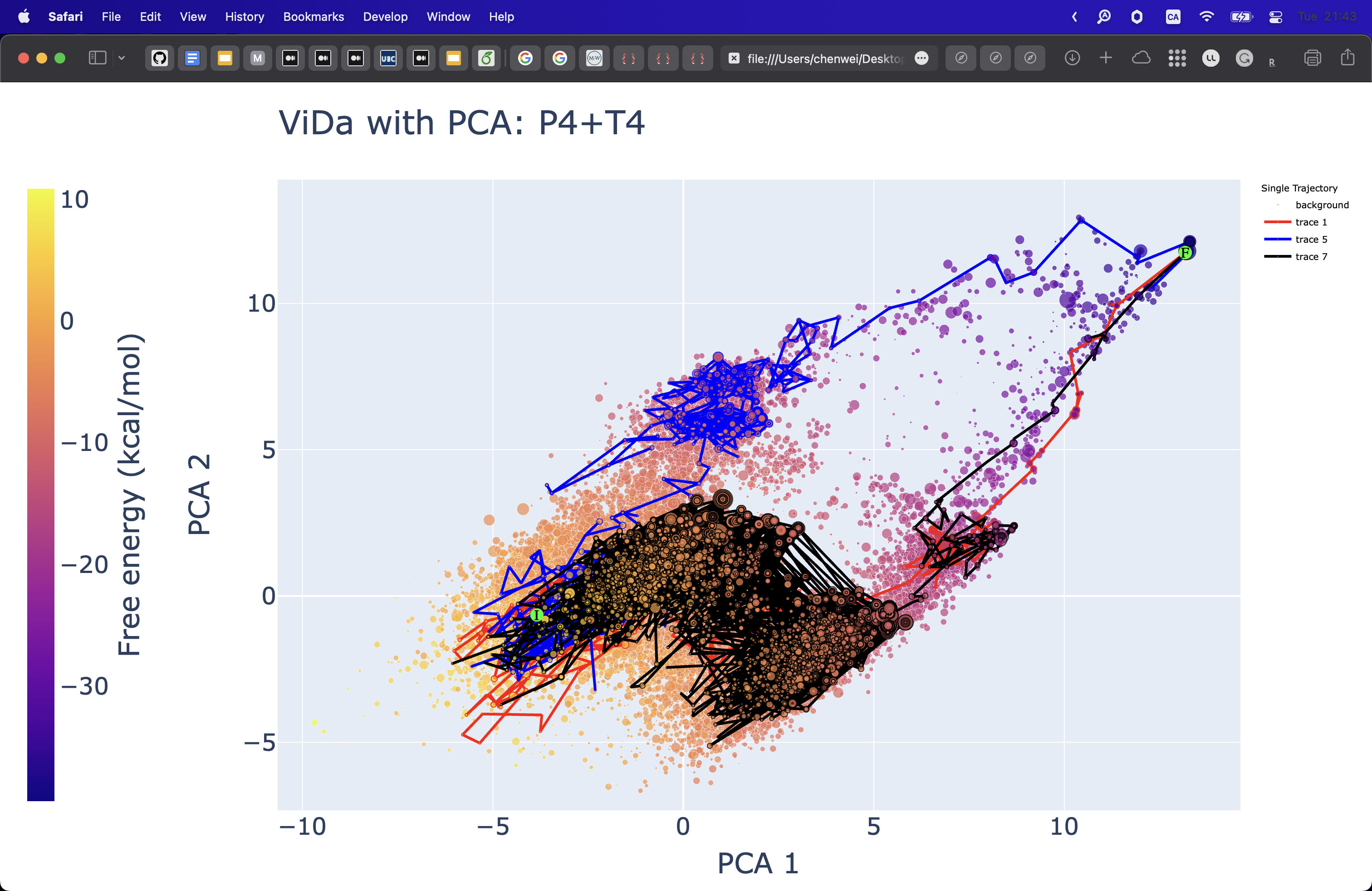}
        \caption{ViDa with PCA: P4+T4 \label{fig:result-d}}
    \end{subfigure}\hfil
    \begin{subfigure}[b]{0.33\textwidth}
        \includegraphics[width=1\linewidth, trim={1cm 0cm 4.5cm 6cm} ,clip] 
        {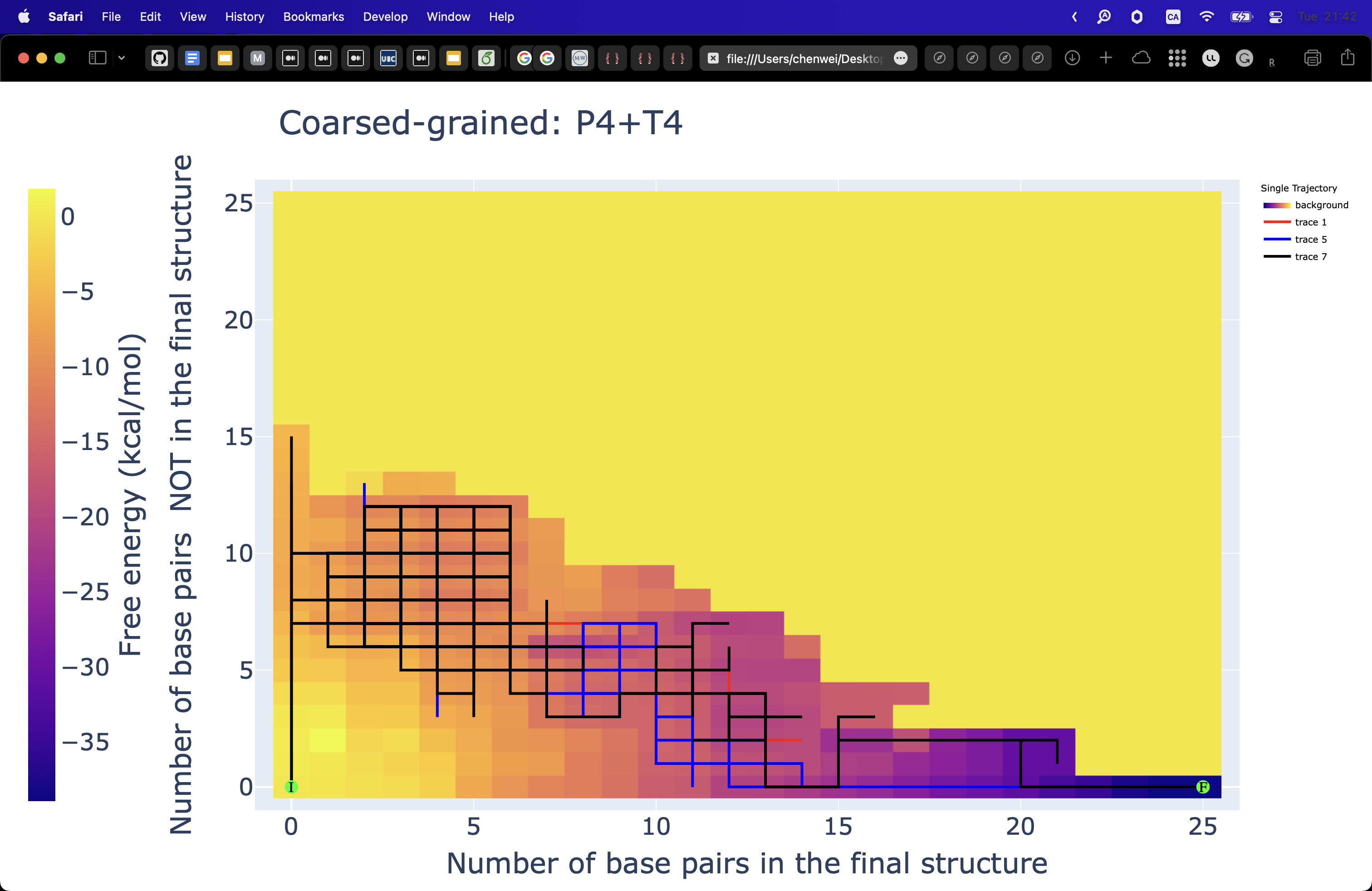}\hfil
        \caption{Coarse-grained: P4+T4 \label{fig:result-cg4}}
    \end{subfigure}\hfil
    \caption{ 
    	 Trajectories laid out on the energy landscapes. Each point in the energy landscape refers to a state with the colour referring to the value of free energy. The initial and final states are indicated by the green circles marked $I$ and $F$, respectively. Each coloured curve represents a trajectory.
    	 \textbf{(a)} The plot made by ViDa with PHATE for sample P0+T0.
    	 \textbf{(b)} The plot made by ViDa with PCA for sample P0+T0.
    	 \textbf{(c)} The plot made by the coarsed-grained method for sample P0+T0. 
    	 We did the same in \textbf{(d)}, \textbf{(e)}, and \textbf{(f)} for sample P4+T4.
    	}
    \label{fig:results}
\end{figure}

\paragraph{ViDa preserves global and local structure in energy landscapes.} \label{finding1}
In the plots of the reaction landscape (Figure \ref{fig:result-a}, \ref{fig:result-b}, \ref{fig:result-c}, and \ref{fig:result-d}), state (secondary structure) distribution in all landscapes follows a high-to-low energy trend from the initial state to the final state, indicating that both PCA and PHATE preserve global structure. PHATE's embedding of sample P0+T0 (Figure \ref{fig:result-a}) separates low energy states into two main branches. The top branch represents states in which most inter-strand base pairs form at the 3$'$ ends of strand P0, e.g., with abbreviated dp structure such as {3$'$-{((((((( [$\ldots$]}-5$'$+3$'$-{[$\ldots$] )))))))}-5$'$} (where ``[$\ldots$]'' denotes a region with a mix of unpaired and paired bases' dp notation). The bottom branch represents states opposite to the top one, such as: {3$'$-{[$\ldots$] (((((((}-5$'$+3$'$-{))))))) [$\ldots$]}-5$'$}.\footnote{We could quickly explore state's structures by using an interactive tool that we implemented, which we do not describe further here.} This suggests that PHATE preserves local structure as well.
More evidence of local structure preservation can be seen in PHATE's embedding of the landscape of P4+T4. High energy states are split into three major branches, labeled as 1, 2, and 3 (Figure \ref{fig:result-c}). Inspection of their secondary structures shows that most of states in branch 1 have the 4-stem hairpin structure (see Table \ref{table:samples} and Figure \ref{fig:4stem}) near either the 3$'$ end of strand P4 or the 5$'$ end of strand T4. In contrast, 3-stem hairpins are dominant in branch 3, while for states in branch 2, hairpins form randomly and are not stable. Although PCA is a linear method, better at preserving global structure, interestingly, we note that two low energy branches are separated by PCA in Figure \ref{fig:result-d}, suggesting PCA can also capture local structure for sample P4+T4. 

\paragraph{ViDa visualizes trajectories in a meaningful way.} \label{finding2}
Laying out the trajectories on the energy landscapes, all trajectories proceed nicely along the branches (Figure \ref{fig:result-a}, \ref{fig:result-c}, and \ref{fig:result-d}). Note that our visualization pipeline does not get actual trajectories as input, but rather just unordered sets of states visited by the trajectories. Therefore, different branches reflect different reaction pathways. For instance, the red trajectory along on the top branch of Figure \ref{fig:result-a} follows the trend in which inter-strand base pairs tend to form from the 3$'$ end of strand P0. Additionally, for all trajectory plots, we did not observe large jumps occurring along the curves, confirming that our embedding tends to cluster similar secondary structures together.

\paragraph{ViDa provides more nuanced understanding of reaction mechanisms.} \label{finding3}
Upon considering the reaction P4+T4 (see Figure \ref{fig:result-c}), our embedding confirms that the 4-stem hairpin at one or both ends of the strands slows the reaction down, as intended by design of the P4+T4 sample. Specifically, many states along the black trajectory have at least one of the 4-stem stable hairpins, providing a barrier to hybridization. In addition, the blue trajectory in branch 3 is also trapped. We determined that this alternative trajectory has 3-stem hairpins (see Figure \ref{fig:3stem}) in a different region of P4+T4, which also contribute to the overall slowness. Occasionally, the strand may fold quickly as the reaction pathway goes through branch 2 in which no stable hairpins exists, such as the red trajectory. Thus, visualizing the blue trajectory can prompt a user to do further analysis of specific kinetic traps that slow down the P4+T4 reaction.

\paragraph{ViDa is more informative than
coarse-grained methods.} \label{finding4}
In the coarse-grained representation of DNA hybridization, as shown in Figure \ref{fig:result-cg0} and \ref{fig:result-cg4}, the X axis refers to the number of base pairs in the final structure, while the Y axis refers to the number of base pairs \emph{not} in the final structure. That is, the grid cell at position $(i,j)$ has $i$ base pairs of the final structure and $j$ base pairs not in the final structure. Accordingly, trajectories for sample P0+T0 and P4+T4 start in grid cell (0,0) and end in grid cell (25,0) (since the strands have length 25). 
With this representation, a coarse-grained grid cell may include states that are unlikely to be on sampled trajectories, and thus are irrelevant, and may also include states with very different free energies, making it difﬁcult to interpret what happens during a DNA reaction. In contrast, ViDa's fine-grained embedding overcomes this limitation. 
Moreover, the coarse-grained trajectories for both samples cross over and overlap with each other so that they are very difficult to discriminate, while ViDa's plots show distinct reaction trajectories, enabling users to interpret reaction mechanisms more straightforwardly and accurately. 

\section{Discussion} \label{suggestion}
In this paper, we introduced ViDa to visualize DNA reaction trajectories, and shown on two specific cases how it allows to comprehend reaction mechanisms. We observed that different choices of dimensionality reduction methods make a difference in showing energy landscapes and corresponding trajectories. We also investigated t-SNE and UMAP (see Figure \ref{fig:results_tsneumap} in Appendix \ref{tsneumap}), and found that these methods do not work well for our data samples. This may result from their default parameter settings, such as perplexity (the number of nearest neighbours) for t-SNE and/or the tightness (minimum distance apart the points) for UMAP. 
Although PHATE works well in the two cases tested here, we note that it sometimes maps distinct states onto the same point in 2D Euclidean space. In contrast, the PCA's mapping from states to points in 2D is one-to-one. 
We also question whether PHATE's separation of the two branches for sample P0+T0 provides meaningful insight, since sample P0+T0 is designed so that no stable hairpins can form the reaction time for trajectories passing through either branch is about the same, and the mechanism is similar (just differing in the direction of base pair ``zipping''). Further work is needed to assess and possibly address these issues with PHATE.

A desirable feature for domain experts in DNA nanotechnology would like, in visualizations of reactions that involve multiple strands, is that the states associated with distinct subsets of ``connected'' strands, i.e., strands that are connected by inter-strand base pairs. For our simple example of DNA hybridization, states with no inter-strand base pairs would ideally be separated from strands with inter-strand base pairs. For reactions involving three strands, such as three-way strand displacement, there should be five distinct groups (one group with no inter-strand base pairs, three groups in which one strand has no base pairs to the other two, which are connected, and one group with all three strands connected). However, ViDa's embeddings do not provide such a separation of groups on our DNA hybridization reaction samples (see Figure \ref{fig:results_pairunpair} in Appendix \ref{connectunconnect}), so this is another important direction for improving the underlying ML models.

Finally, there are many additional opportunities to further evaluate and build on the current basic capabilities of ViDa, and improve the underlying ML tools. Our next steps are to assess ViDa's visualizations of strand displacement, given the importance of this three-stranded reaction. We are currently interested in assessing whether for this and other more complex reactions, it is necessary to train graph embedding methods on trajectories, rather than simply sets of visited states.

\begin{ack}
This work has been supported by NSERC Discovery Grant. The authors thank Boyan Beronov, Jordan Lovrod, and Erik Winfree for their insightful feedback. 
\end{ack}

\newpage

\bibliographystyle{abbrv}
\bibliography{bibli}

\appendix

\section*{Appendices}

\section{The Multistrand simulator} \label{multistrand}
Multistrand is a coarse-grained model designed to simulate thermodynamic and kinetic process for various DNA and/or RNA-strand interactions ignoring formation of pseudoknotted structures. As the name suggests, Multistrand is able to handle complex systems with multiple strands. The simulator uses a Gillespie sampling approach, rather than representing the entire state space of secondary structures explicitly, since the state space could have size exponential in the length of the strands. Transitions between neighbour states are based on elementary steps, i.e., a single base pair forming or breaking. This kinetic model satisfies the detailed-balance condition so that it will reach the equilibrium state distribution given sufficient simulation time, which is in line with thermodynamic predictions made by both NUPACK \cite{nupack} or Vienna RNA \cite{vienna} models.

There are several output modes in the Multistrand simulator for users' different purposes. The simplest one is trajectory mode which was used in this work. With this mode, a specific trajectory is produced through the secondary structure state space. The outputs include a sequence of secondary structures represented by the dp notation, the reaction simulation time (in terms of sampled trajectory time, as opposed to the wall-clock time) and the corresponding free energy of the secondary structure.

\section{Dot-parenthesis notation} \label{dp}
Dot-parenthesis (dp) notation is a simple way to represent a secondary structure of DNA or RNA. Each character represents a base. Dots indicate unpaired bases and matching parentheses indicate paired bases. The number of open and closed parentheses is always equal.
The symbol “+” in the dp notation separates strands. For example, in the dp notation  3$'$-{...(((...}-5$'$+3$'$-{...)))...}-5$'$ for the secondary structure of two DNA strands $A$ (3$'$-TGACGATCA-5$'$) and $\bar{A}$ (3$'$-TGATCGTCA-5$'$), the left part of the ``+'' sign corresponds to strand $A$ and the right part corresponds to strand $\bar{A}$. Three open parentheses indicate that the bases ``CGA'' in strand $A$ are paired with the bases ``TCG'' in strand $\bar{A}$ which are represented by three closed parentheses.

\section{Hairpin schematic representations} \label{hairpin}

\renewcommand{\thefigure}{A\arabic{figure}}
\setcounter{figure}{0}

\begin{figure}[h]
    \centering
    \begin{subfigure}[b]{0.49\textwidth}
        \includegraphics[width=1\linewidth, trim={0cm 0cm 11.5cm 0cm} ,clip] 
        {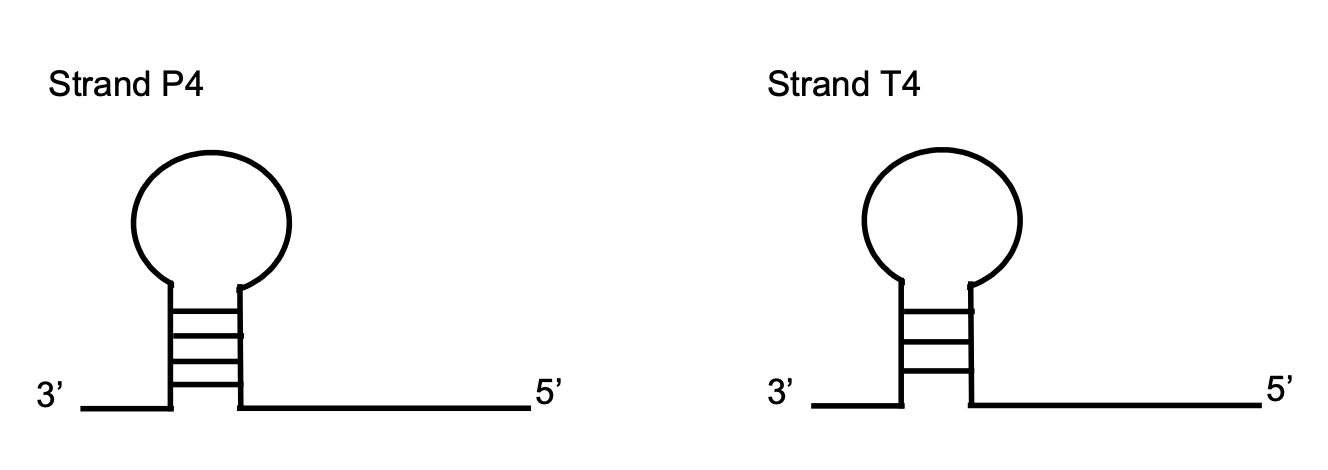}\hfil
        \caption{4-stem hairpin \label{fig:4stem}}
    \end{subfigure}
    \begin{subfigure}[b]{0.5\textwidth}
        \includegraphics[width=1\linewidth, trim={11.5cm 0cm 0cm 0cm} ,clip]
        {Figures/hairpin_schematic.png}
        \caption{3-stem hairpin \label{fig:3stem}}
    \end{subfigure}\hfil
    \caption{ 
    Schematic representations of hairpin structures.
    \textbf{(a)} A 4-stem hairpin (hairpin with four paired bases) formed near 3$'$ end of strand P4.
    \textbf{(b)} A 3-stem hairpin (hairpin with three paired bases) formed near 3$'$ end of strand T4.
    	}
    \label{fig:hairpins}
\end{figure}

\clearpage

\section{ViDa with t-SNE and UMAP} \label{tsneumap}
The t-SNE paramters were set that the perplexity was 30 and the learning rate was 200. For UMAP, the size of the local neighborhood was set to 15 and the minimum distance was set to 0.1.

\begin{figure}[h]
    \centering
    \begin{subfigure}[b]{0.49\textwidth}
        \includegraphics[width=1\linewidth, trim={1cm 0cm 4.5cm 6cm} ,clip] 
        {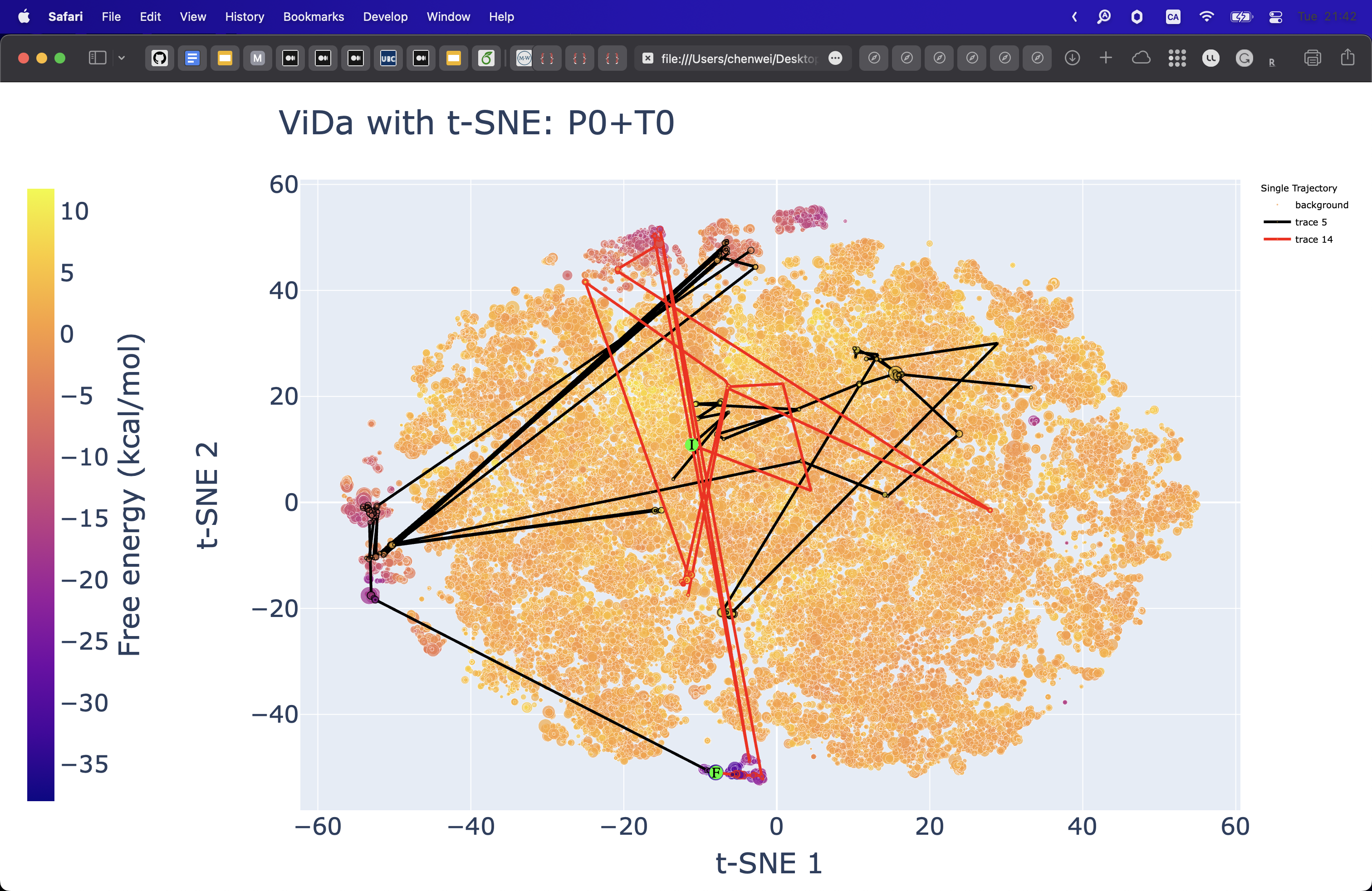}\hfil
        \caption{ViDa with t-SNE: P0+T0 \label{fig:result-tsne0}}
    \end{subfigure}
    \begin{subfigure}[b]{0.5\textwidth}
        \includegraphics[width=1\linewidth, trim={1cm 0cm 4.5cm 6cm} ,clip]
        {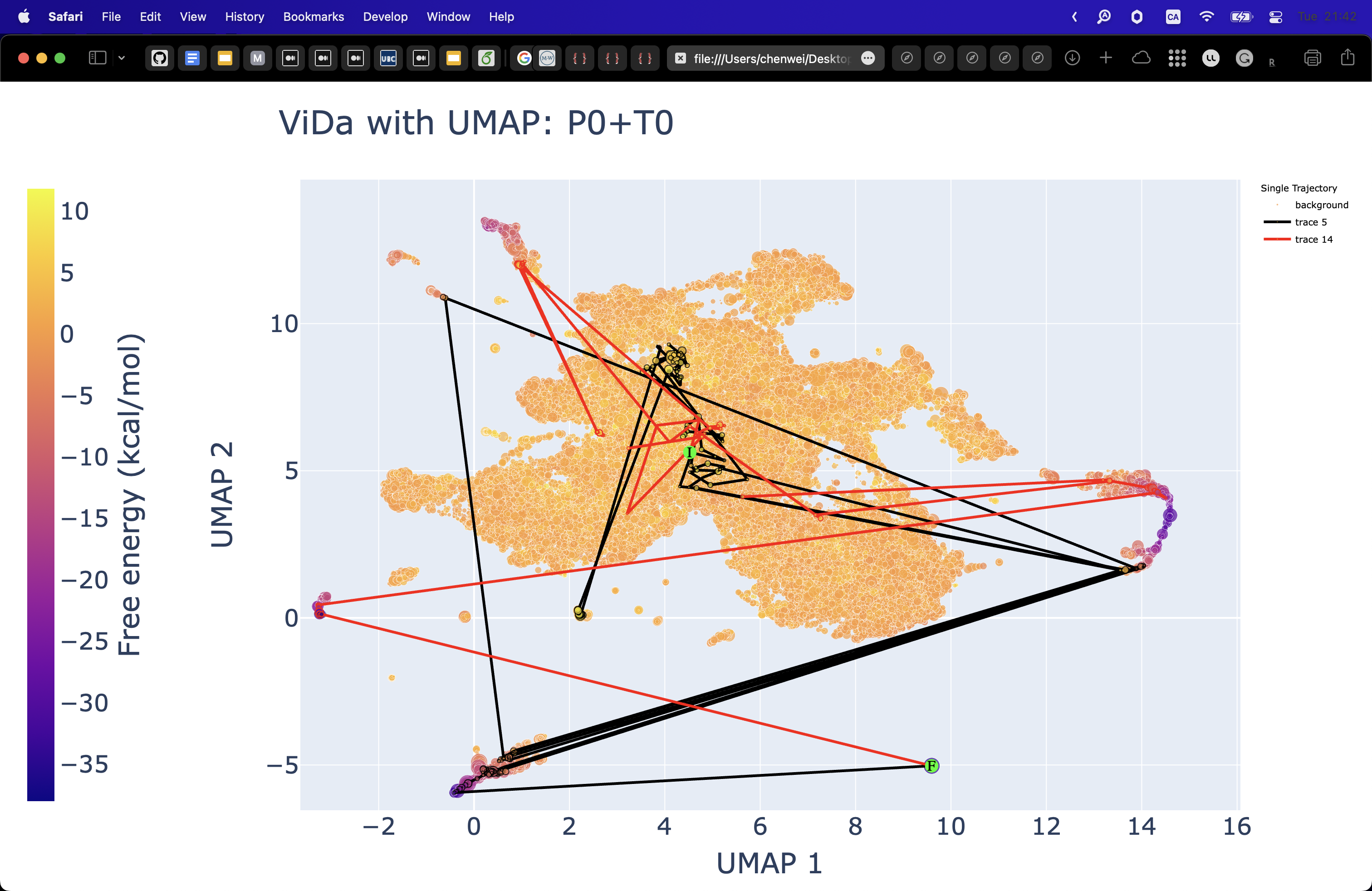}
        \caption{ViDa with UMAP: P0+T0 \label{fig:result-umap0}}
    \end{subfigure}\hfil
    \begin{subfigure}[b]{0.49\textwidth}
        \includegraphics[width=1\linewidth, trim={1cm 0cm 4.5cm 6cm} ,clip] 
        {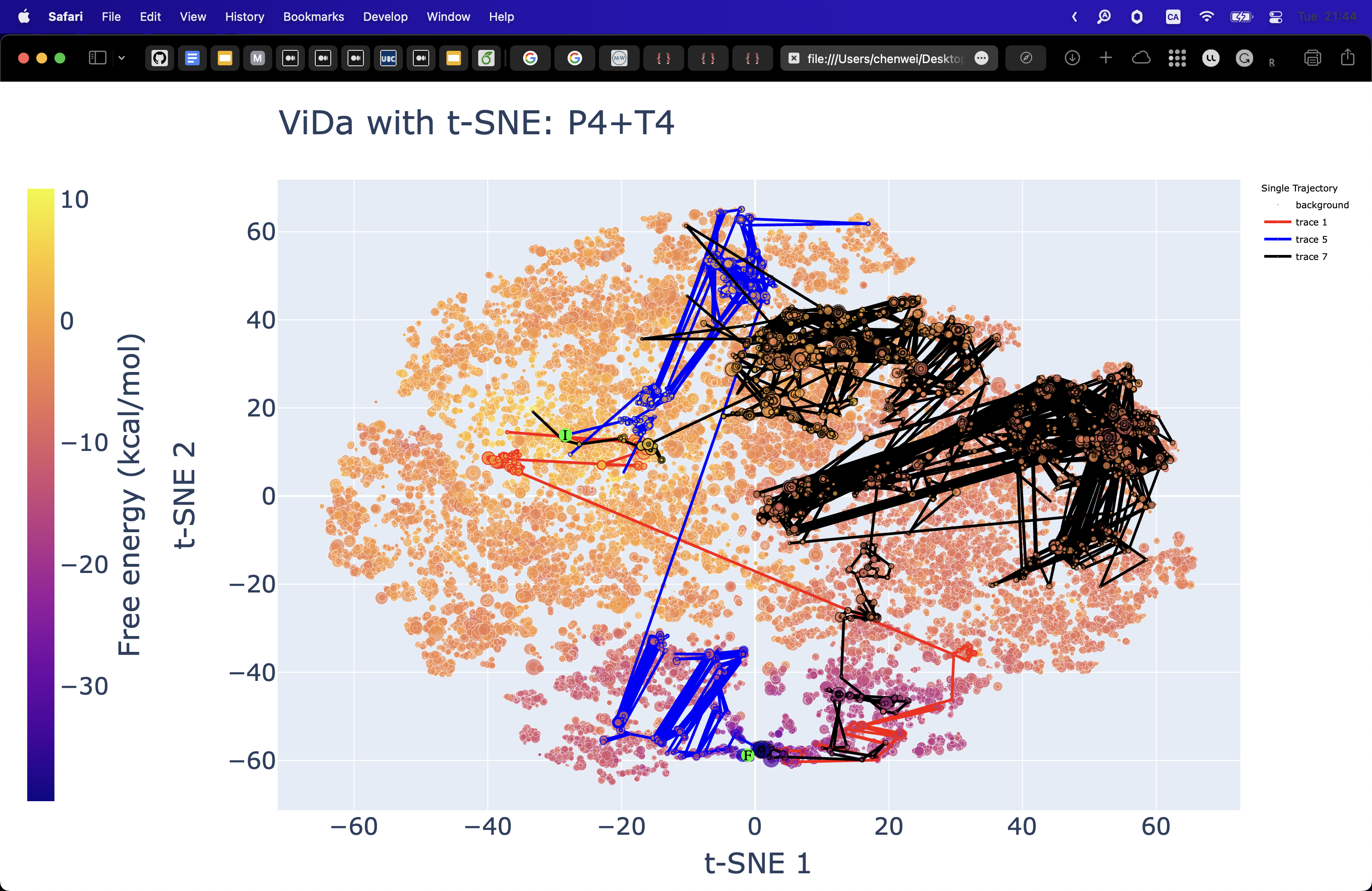}\hfil
        \caption{ViDa with t-SNE: P4+T4 \label{fig:result-tsne4}}
    \end{subfigure}
    \begin{subfigure}[b]{0.5\textwidth}
        \includegraphics[width=1\linewidth, trim={1cm 0cm 4.5cm 6cm} ,clip]
        {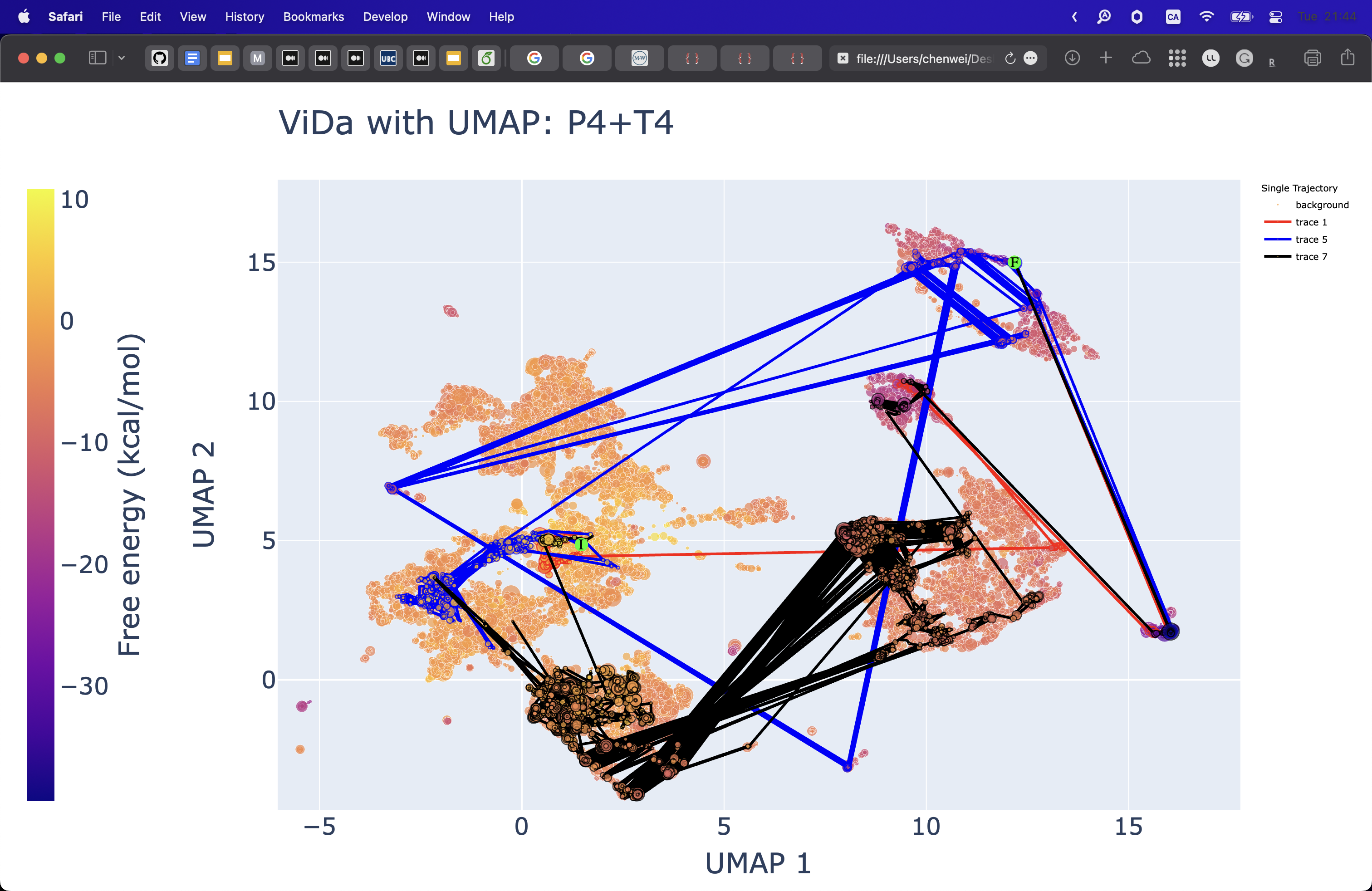}
        \caption{ViDa with UMAP: P4+T4 \label{fig:result-umap4}}
    \end{subfigure}\hfil
    \caption{ 
    Trajectories laid out on the energy landscapes. Each coloured curve represents a trajectory. The initial and final states are indicated by the green circles marked $I$ and $F$, respectively.
    \textbf{(a)} The plot made by ViDa with t-SNE for sample P0+T0.
    \textbf{(b)} The plot made by ViDa with UMAP for sample P0+T0.
    We did the same in \textbf{(c)} and \textbf{(d)} for sample P4+T4.
    	}
    \label{fig:results_tsneumap}
\end{figure}

Figure \ref{fig:results_tsneumap} shows energy landscapes and trajectory plots produced by ViDa with t-SNE and UMAP. It can be seen that neither method preserves global structure well. The distribution of clusters does not follow the energy trend, and trajectories ``jump'' around the landscapes. In future work, we will see if improved hyperparameters can help avert these problems. 

\clearpage

\section{ViDa's energy landscapes for connected and unconnected secondary structures} \label{connectunconnect}

\begin{figure}[h]
    \centering
    \begin{subfigure}[b]{0.49\textwidth}
        \includegraphics[width=1\linewidth, trim={1cm 0cm 2cm 6.5cm} ,clip] 
        {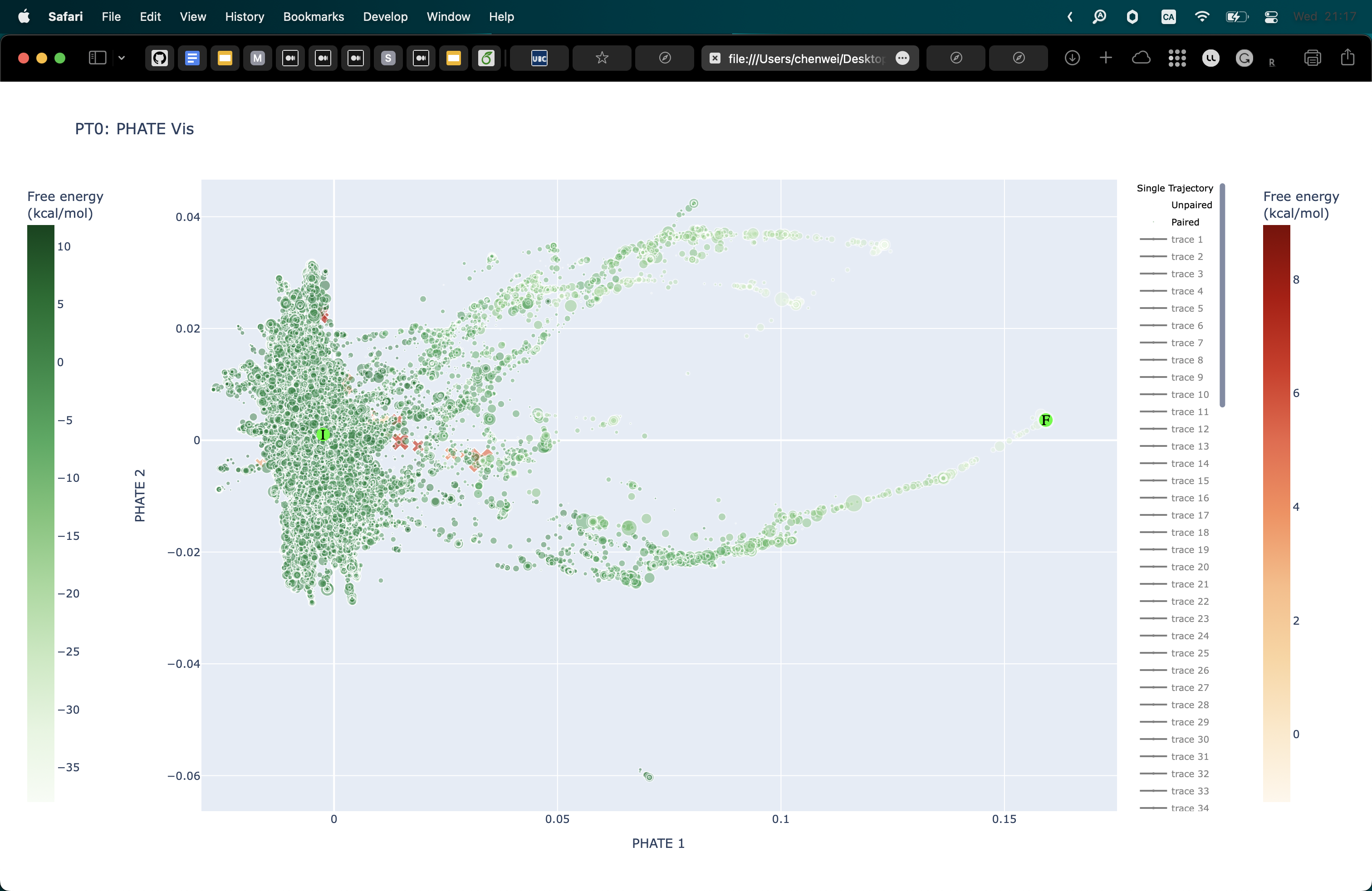}\hfil
        \caption{ViDa with PHATE: P0+T0 \label{fig:result-pairphate0}}
    \end{subfigure}
    \begin{subfigure}[b]{0.5\textwidth}
        \includegraphics[width=1\linewidth, trim={1cm 0cm 2cm 6.5cm} ,clip]
        {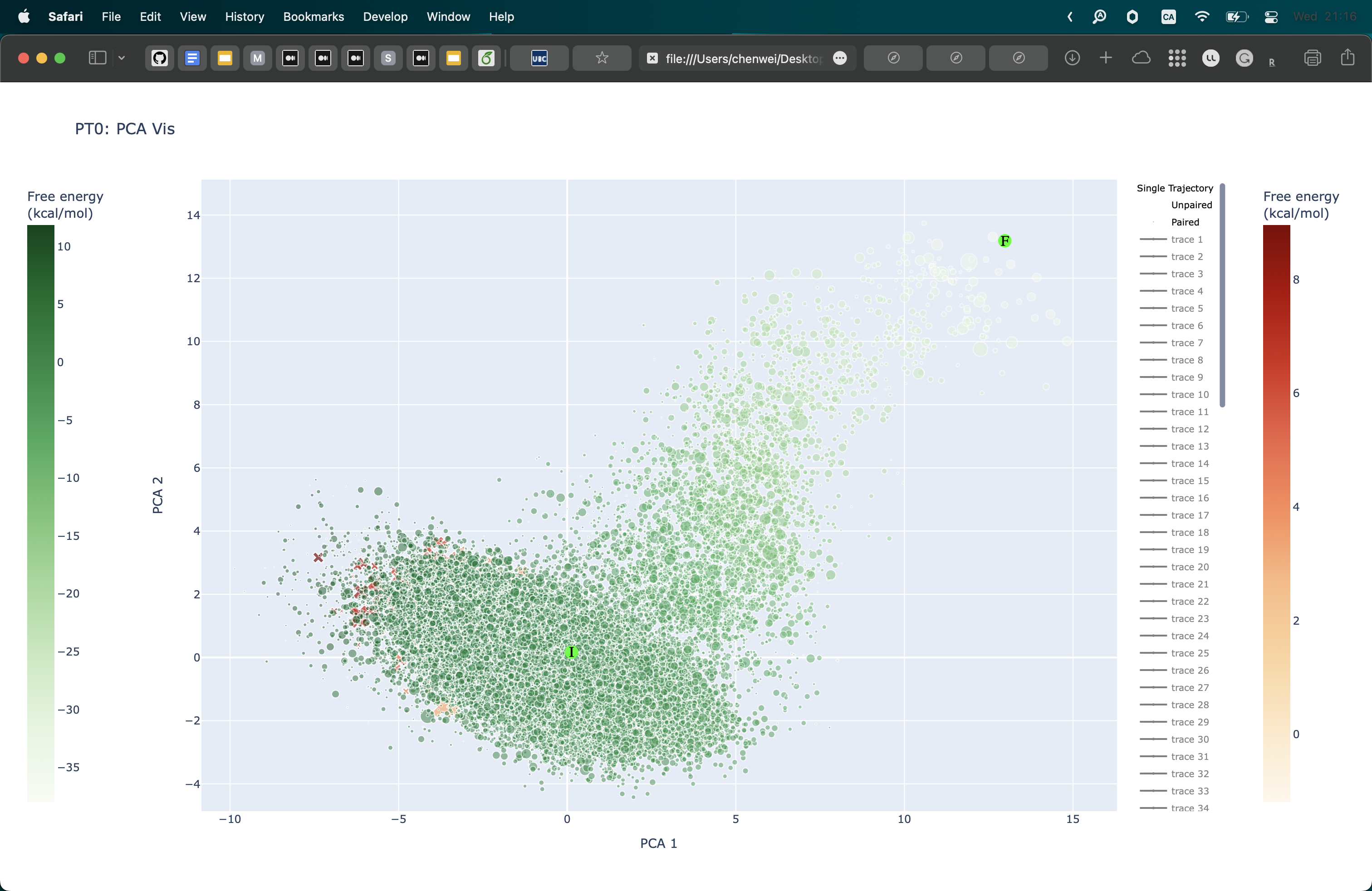}
        \caption{ViDa with PCA: P0+T0 \label{fig:result-pairpca0}}
    \end{subfigure}\hfil
    \begin{subfigure}[b]{0.49\textwidth}
        \includegraphics[width=1\linewidth, trim={1cm 0cm 2cm 6.5cm} ,clip] 
        {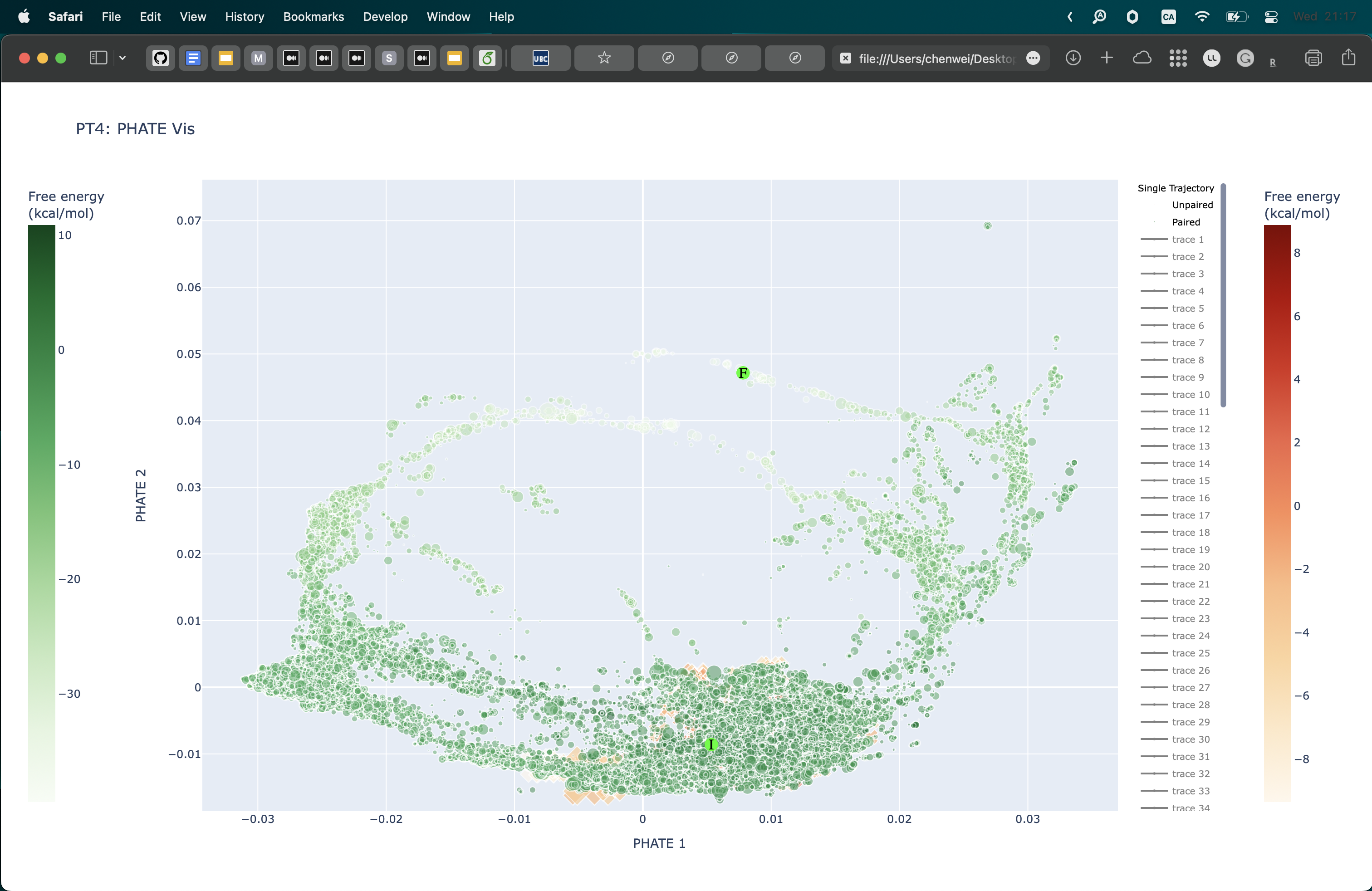}\hfil
        \caption{ViDa with PHATE: P4+T4 \label{fig:result-phatepair4}}
    \end{subfigure}
    \begin{subfigure}[b]{0.5\textwidth}
        \includegraphics[width=1\linewidth, trim={1cm 0cm 2cm 6.5cm} ,clip]
        {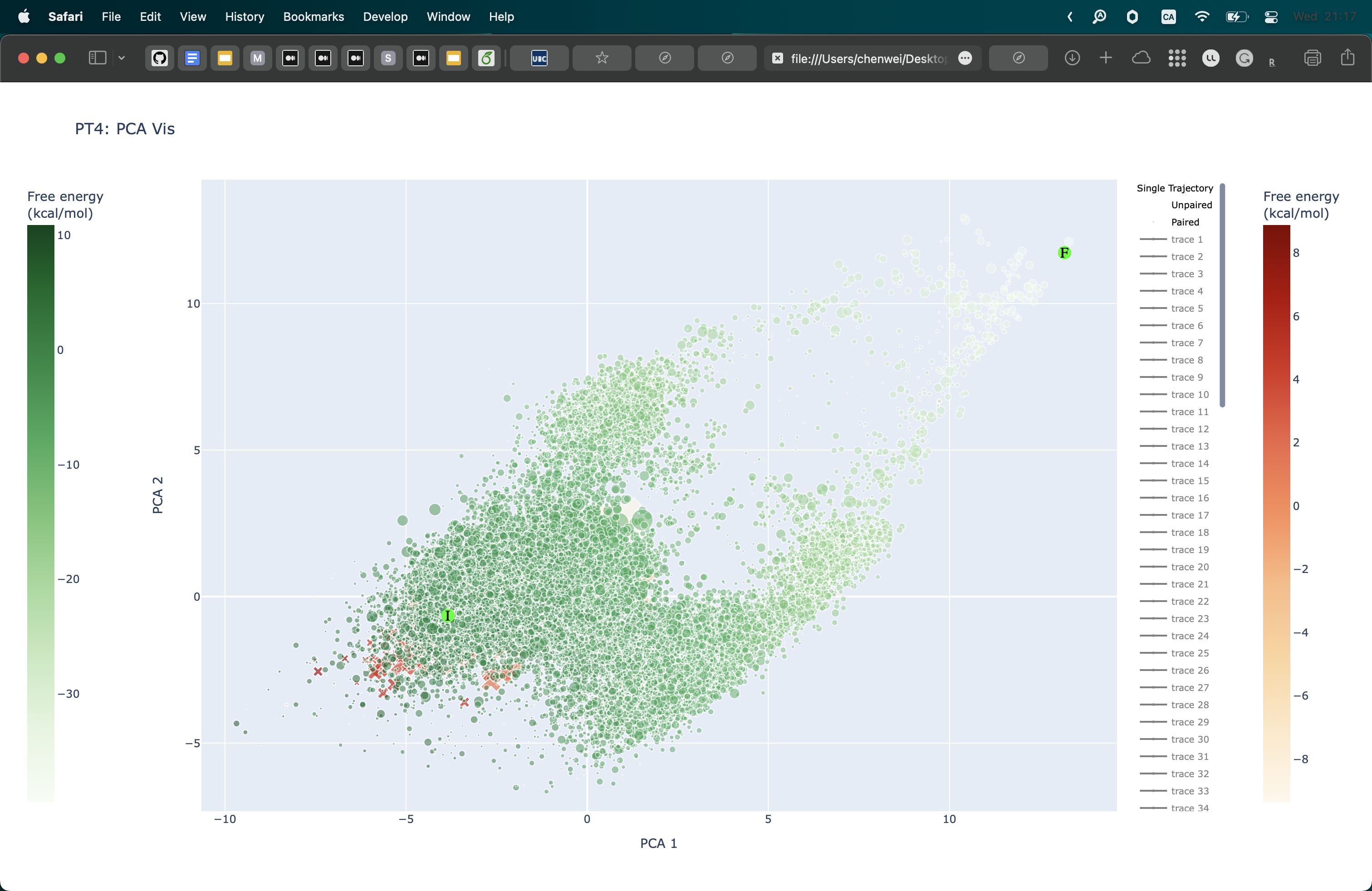}
        \caption{ViDa with PCA: P4+T4 \label{fig:result-pairpca4}}
    \end{subfigure}\hfil
    \caption{ 
    Energy landscapes for connected and unconnected secondary structures. The circles of green colour refer to connected structures. The crosses of red colour refer to unconnected structures. The brightness of colour represents the value of free energy. Lighter colour refers to lower energy, and vice versa. The initial and final states are indicated by the green circles marked $I$ and $F$, respectively.
    \textbf{(a)} The plot made by ViDa with PHATE for sample P0+T0.
    \textbf{(b)} The plot made by ViDa with PCA for sample P0+T0.
    We did the same in \textbf{(c)} and \textbf{(d)} for sample P4+T4.
    	}
    \label{fig:results_pairunpair}
\end{figure}

In Figure \ref{fig:results_pairunpair}, connected secondary structures (i.e., those with at least one inter-strand base pair)  are depicted using circles, while structures with two single-stranded components are depicted using crosses. We can see that circles cover most of crosses, which is unfortunate. Domain experts would appreciate a dimensionality reduction method that distinguishes between states with inter-strand base pairs from those with no such pairs, keeping them separate from each other. Generalizing our methods when there are multiple interacting strands, and thus many different possible connected components involving different subsets of the strands, presents an interesting research challenge.

\end{document}